\documentclass[12pt]{article}
\usepackage{amsmath,amssymb,graphicx,color,leqno,hyperref,url,float}

\usepackage[a4paper, total={6.25in, 8.75in}]{geometry}

\usepackage{setspace}

\usepackage{enumitem}
\usepackage{lscape}

\newcommand{\mbf}[1]{\mbox{\boldmath $#1$}}
\newcommand{\ba}{{\mbf \beta}}

\newtheorem{lem}{Lemma}[section]
\newtheorem{rem}{Remark}[section]
\newtheorem{thm}{Theorem}[section]

\numberwithin{thm}{section}

\numberwithin{equation}{section}

\newcommand{\cL}{{\cal L}}
\newcommand{\cM}{{\cal M}}

\newcommand{\cS}{{\cal S}}

\newcommand{\cU}{{\cal U}}

\def\ba{\begin{array}}
\def\bc{\begin{center}}
\def\bd{\begin{description}}
\def\be{\begin{enumerate}}
\def\ea{\end{array}}
\def\ec{\end{center}}
\def\ed{\end{description}}
\def\edt{\end{document}}
\def\ee{\end{enumerate}}
\def\ben{\begin{equation}}
\def\benn{\begin{equation*}}
\def\een{\end{equation}}
\def\eenn{\end{equation*}}
\def\benr{\begin{eqnarray}}
\def\eenr{\end{eqnarray}}
\def\benrr{\begin{eqnarray*}}
\def\eenrr{\end{eqnarray*}}

\def\del{\delta}
\def\edt{\end{document}}

\def\h{\hat}

\def\hs{\hskip}
\def\iny{\infty}

\def\la{\lambda}
\def\lel{\label}

\def\mb{\mbox}
\def\noi{\noindent}

\def\nn{\nonumber}

\def\Om{\Omega}
\def\om{\omega}

\def\r{\ref}

\def\si{\sigma}

\def\Si{\Sigma}

\def\vs{\vskip}

\def\R{{\mathbb R}}

\begin{document}

\doublespacing

\bc
{\bf \Large Analysis of High Dimensional Compositional Data Containing Structural Zeros with Applications to Microbiome Data}\\[0.5cm]
Abhishek Kaul$\footnote{Corresponding author. E-mail: abhishek.kaul@nih.gov, Address: 111 T.W. Alexander dr., Rm A-385, RTP, NC 27709. },$ Ori Davidov$^{2}$ and Shyamal D. Peddada$^{1}$\\[0.25cm]
	$^{1}$Biostatistics and Computational Biology Branch,\\ National Institute of Environmental Health Sciences, NC, USA\\ $^{2}$Department of Statistics, University of Haifa, Haifa, Israel
\ec
\vs .1in

\begin{abstract}
This paper is motivated by the recent interest in the analysis of high dimensional microbiome data. A key feature of this data is the presence of `structural zeros' which are microbes missing from an observation vector due to an underlying biological process and not due to error in measurement. Typical notions of missingness are insufficient to model these structural zeros. We define a general framework which allows for structural zeros in the model and propose methods of estimating sparse high dimensional covariance and precision matrices under this setup. We establish error bounds in the spectral and frobenius norms for the proposed estimators and empirically support them with a simulation study. We also apply the proposed methodology to the global human gut microbiome data of Yatsunenko (2012).
\end{abstract}
\noi {\bf Keywords:} Microbiome data, High Dimension, Classification, Sparsity, Missing data.
%

\section{Introduction}
With the advancement of high throughput technologies, it is {now} common to encounter high dimensional data {with the} number of parameters ($d$), often far exceeding the sample size ($n$). {In this high dimensional setting it is often of interest to investigate relationships among thousands of variables.}

This paper is motivated by the recent surge in interest to understand the effects of microbiome on our {external and internal environment and also on public health}. {For example, it is often of interest to understand the relationships among various bacterial populations and how such relationships may affect health outcomes. In some cases it may also be of interest in identifying microbial biomarkers which can classify subjects into two different populations using microbiome data. A detailed review of recent literature on this topic is provided by (cf  Clemente et. al., 2012)}

In order to address such scientific questions, one needs to first estimate the covariance matrix $(\boldsymbol\Si)$ or its inverse, the precision matrix $(\boldsymbol\Om=\boldsymbol\Si^{-1}).$ Estimation of $\boldsymbol\Si$ and $\boldsymbol\Om$, when the dimension exceeds the sample size, i.e. $n \le d$ has been discussed extensively in the literature.  The existing literature can be broadly classified into two categories, the first approach involves estimation of the precision matrix by exploiting its natural sparsity in comparison to the covariance matrix [cf.. Friedman, Hastie and Tibshirani, 2007, Cai, Liu and Luo, 2011, and Rothman, Bickel, Levina and Zhu, 2008]. A limitation of this approach is that it does not apply to low rank matrices $\boldsymbol\Si$ since the precision matrix does not exist in this case. The second popular approach is to estimate the $\boldsymbol\Si$  by assuming that $\boldsymbol\Si$ is itself sparse. One of several methods for this purpose is to threshold each element of the sample covariance matrix [Bickel and Levina 2008, and Rothman, Levina and Zhu, 2009].

All papers mentioned above assume the availability of independent and identically distributed (i.i.d) copies of the vector ${\bf X}=(X_1,X_2,...,X_d)^T$ whose distribution is Gaussian or more generally sub-Gaussian with $\boldsymbol\mu$ and $\boldsymbol\Si$ as the $d$ dimensional mean vector and covariance matrix respectively.  Note that a real valued random variable $X_1$ is said to be sub-Gaussian if there exists a $b>0$ such that for every $t\in \R,$ one has $Ee^{tX_1}\le e^{b^2t^2/2}.$

In contrast to typical high dimensional data, not all variables (i.e. microbes) are observed in {a microbial expression sample.} Thus if ${\bf X}$ represents a $d$ dimensional vector of abundances of $d$ taxa in a specimen obtained from an ecosystem, then not all components of ${\bf X}$ may be observed. {We refer to this missingness as structural zeros and it is due to the underlying biology and not not due to error in measurement or values below the minimum detection level. For example, it is known that the bacterial genus {\it Bacteroides} is prevalent in the human gut when the associated diet is high protein/fat diet, whereas it may be completely absent otherwise, i.e. carbohydrate rich diet.} The total abundance of such bacteria are coded as 0 counts in the observational vector ${\bf X}$.

The missing structure required to model structural zeros is more general than typical notions of missingness in the literature. More precisely, in the classical notions of missingness, such as missing completely at random (MCAR) or missing at random (MAR), it is assumed that in place of ${\bf X}$ we observe a surrogate vector ${\bf U}={\bf X}\oplus{\bf W},$ where $\oplus$ represents a component-wise product and ${\bf W}$ is a $d$-dimensional vector of independent Bernoulli random variables. In effect, not all components of ${\bf X}$ are observed in ${\bf U}$. For example, ${\bf U}=(0,0,X_3,..,X_p)^{T},$ corresponds to the case where  the first two components of ${\bf X}=(X_1,...,X_d)^T$ are not observed in ${\bf U}$ with ${\bf W}=(0,0,1,...,1)^T.$ In this example, although $X_1$ and $X_2$ are absent in ${\bf U},$ they still influence the distribution of the remaining components $X_3,..,X_p$ through the underlying dependence structure of $\boldsymbol\Si$ and are only hidden by the corresponding multiplicative Bernoulli noise vector ${\bf W}.$ In contrast, for the case of structural zeros the observed vector itself is ${\bf X}=(0,0,X_3..,X_p),$ i.e., the first two components are truly absent from the observation and thus the missing components should not influence the distribution of the remaining components.

{In this paper we define a general framework which allows for structural zeros in the model and discuss consistent methods of estimating sparse high dimensional covariance and precision matrices under this setup. We establish consistency in estimation of the proposed methodology and empirically support it with a simulation study. We also apply our methodology to analyse the global human gut microbiome data of Yatsunenko et. al. 2012. Estimation of covariance and precision matrices in the traditional missing values setting has also been discussed in the literature [cf. Loh and Wainwright , 2012) and Lounici, 2012]. As shall become apparent in the following, our model allows for a more general notion of missingness while assuming weaker conditions in comparison to typical notions of missingness.}

\section{Notations and Framework} \lel{notations}

Throughout the paper, for any $l\times m$ matrix ${\bf A}=[a_{ij}]$ define the $\ell_0,$ $\ell_1,$ $Sup,$ $Spectral$ and $Frobenius$ norms as $\|{\bf A}\|_0=\textnormal{Card}\{ij\,:\, A_{ij}\ne 0\},$ $\|{\bf A}\|_1=\sum_{i,j}|a_{ij}|,$ $\|{\bf A}\|_{\iny}=\max_{i,j}|a_{ij}|,$ $\|{\bf A}\|_2=\sup_{||x||_2\le 1}||Ax||_2$ and $\|{\bf A}\|_{F}=\sqrt{\sum_{i,j}a_{ij}^2}$, respectively. Also ${\bf A}\succ 0$ indicates the matrix ${\bf A}$ is positive definite. We use $c_0,$ $c_1$ and $c_2$ as generic constants which may change according to the context. For any set of indices $S,$ its cardinality is denoted by $|S|.$ For a subset $A\subseteq \{1,2,\cdots,d\}$, ${\bf b}_A$ denote the vector of components of ${\bf b}$ with indices in $A.$ Also a $p\times p$ matrix $\boldsymbol\Si$  is partitioned as
\benr\lel{sigpar}
\boldsymbol\Si = \left(\begin{matrix} \boldsymbol\Si_{AA} & \boldsymbol\Si_{AA^c}\\ \boldsymbol\Si _{A^cA} & \boldsymbol\Si_{A^cA^c}
\end{matrix}\right),\qquad  \mb{where $A^c$ denote the compliment set of A.}
\eenr

We begin by describing a framework that characterizes structural zeros. As briefly stated in the Introduction, these structural zeros represent components that are biologically absent in the specimen. Hence, intuitively the framework should allow for the distribution of the specimen to be completely determined by only the observed components. Restating this statistically, the distribution of an observation should be characterized conditional to the missing structure for each $1\le i\le n$. Hence we first define the missing structure.

Let the sample space ${\cal S}$ of possible configurations of missing components in a given sample be as follows.
\benr\lel{label}
\cS=\begin{cases}
      (1,\ldots, 1), \\
      (0,1,\ldots, 1), (1,0,\ldots, 1), \ldots,  (1,\ldots, 1,0) \\
      (0,0,1,...1), (0,1,0..,1), \ldots,   (1,\ldots, 0,0)\\
      .\\
      .\\
      (0,0,\ldots, 1), (0,0,\ldots, 1,0), \ldots, (1,0,...0)
\end{cases}
\eenr
Here $0,1$ correspond to the cases where a component is unobserved or observed in the sample respectively. We shall represent each of the above $2^d -1$ events of the sample space by Configuration $(j)$, $j = 1, 2, \ldots, 2^d -1$, in the order written in (\r{label}). For example, Configuration ($1$) is the case where all components are observed and Configuration $(2^d-1)$ corresponds to the configuration where only the first component is observed. For each sample $i$, $1\le i\le n,$ we assume that the missing structure is generated by independent random variables ${\bf M}_i,$ $1\le i \le n,$ with sample space described in (\r{label}).

{In many applications, it may be unreasonable to assume that the missingness is generated by identically distributed r.v.'s. The distriubtion function may be influenced by factors or covariates such as geographical location, age, race and gender of the subject.} To allow for this flexibility, let ${\bf z_i},$ $1\le i\le n$ be $q$-dimensional vectors of non-random covariates which can possibly influence the distribution of the missingness, more precisely, define the distribution of the random variables ${\bf M}_i,$ $1\le i\le n$ by,
\benr\lel{delta}
P\Big({\bf M}_i\,\, \textnormal {is in Configuration}\,\, (j)\,\,\Big) =\delta_{(j)}({\bf z}_i),\quad 0\le \delta_{(j)}({\bf z}_i)\le 1,\quad \,1\le j\le 2^d-1.
\eenr
{This feature of allowing the distribution to be influenced by factors or covariates while preserving independence is reminiscent of the MAR structure of missingness.} We now proceed to define the conditional distribution of the observed components of a specimen.

Let $\boldsymbol\mu=(\mu^{1},...,\mu^{d})^{T},$ $\mu^{k}\in\R$  and $\boldsymbol \Si=[\si_{ij}]_{d\times d}$ be a d-dimensional vector and symmetric matrix respectively. For a subject $i,$ with missing configuration given by the random variable ${\bf M}_i$, we denote the observed components by the index set
\benr
A_i=\{j,\, M_{ij}=1\}.
\eenr
Note that the index set $A_i$ is a random set which is determined by the r.v. ${\bf M}_i.$ Now assume that conditioned on ${\bf M}_i,$ the components of ${\bf X}_i$  with indices in the index set $A_i$ jointly follow a Gaussian distribution with mean and covariance being the corresponding sub-vector of $\boldsymbol\mu_i$ and sub-matrix of $\boldsymbol \Si $ respectively, i.e., for any ${\bf x}\in \R^{d},$
\benr\lel{yid}
P\Big({\bf X}_{A_i} \le {\bf x}_{A_i}\Big| {\bf M}_i\Big)=\Phi_{A_i}({\bf x}_{A_i}),
\eenr
where $\Phi_{A_i}$ represents the Gaussian distribution function with mean $\boldsymbol\mu_{A_i}$ and covariance matrix $\boldsymbol\Si_{A_iA_i}.$ For example,  let ${\bf M}_i =(1,1,0,...,0)$, then the observed vector is ${\bf X}_i=(X_{i1},X_{i2},0...,0)$ with the conditional distribution of the observed components as $P\Big(X_{i1}\le x_{i1}, X_{i2}\le x_{i2}\Big| {\bf M_i}\Big)=\Phi(x_{i1},x_{i2}).$

For $1\le l, m\le d$ let
\benr\lel{nlm}
n(l)= \{ i\,\,:\,\, l\in A_i,\, 1\le i\le n \}, \quad {\rm and}\quad n(l,m)= \{ i\,\,:\, l,m \in A_i\,,1\le i\le n \}\nn
\eenr
be the number of subjects where $l^{th}$ component is observed and the number of subjects where the $l^{th}$ and $m^{th}$ components are observed respectively. Note that these are random quantities.

For a given subject $i = 1, 2, \ldots, n$, with covariate vector ${\bf z}_i$, and for $1\le l,m\le d,$ define
\benr
C_{{\bf z}_i}(l)&=&\big\{1\le j\le 2^d-1,\,\,\textnormal{component $l$ is present in Configuration $(j)$}\nn\\
&& \hs 1.9in \textnormal {with covariate ${\bf z}_i$}\big\},\nn\\
C_{{\bf z}_i}(l,m)&=&\{1\le j\le 2^d-1,\,\,\textnormal{components $l$ and $m$ are present in Configuration $(j)$}\nn\\
&& \hs 1.75in \textnormal {with covariate ${\bf z}_i$}\big\}
\eenr
In the sequel we make the following additional assumption over the missing structure.
\begin{description}
\item[{\bf (A1)}] There exists a constant $\delta_{\min}>0$ such that for any $1\le l,\,m\le d,$
\benr
\textnormal {(i)}\,\,\frac{1}{n}\sum_{i=1}^{n}\sum_{j\in C_{{\bf z}_i} (l)}\delta_{(j)}({\bf z}_i)=\delta{(l)}>\delta_{\min}\quad\textnormal{(ii)}\,\,\frac{1}{n}\sum_{i=1}^{n}\sum_{j\in C_{{\bf z}_i}(l,m)}\delta_{(j)}({\bf z}_i)=\delta(l,m)>\delta_{\min}.\nn
\eenr
\end{description}
Note that {\bf (A1)} is a mild assumption on the missing structure. When there are no covariates, (i) reduces to $\sum_{j \in C(l)}\delta_{(j)}>\delta_{\min},$ and (ii) reduces to  $\sum_{j \in C(l,m)}\delta_{(j)}>\delta_{\min}$. Thus in this case, Assumption {\bf (A1)} requires that each component is present in an observational vector with a nonzero probability and that every pair of components are present in each observational vector with a nonzero probability.

\section{Estimation of the Covariance and Precision Matrices}\lel{est}

In this section we derive the theoretical properties of two methodologies, a generalised thresholding procedure to estimate the covariance matrix $\boldsymbol\Si$ and a $\ell_1$ minimisation approach to estimate the precision matrix $\boldsymbol\Om.$ We shall derive these properties under the structural zero's setup while allowing the dimension of the observed vector to increase exponentially with the sample size. The consistency results to follow later in this section shall hold for the following class of approximately sparse matrices.

\begin{description}
\item[{\bf (A2)}] We assume that the covariance and precision matrices belong to the following classes of matrices respectively:
\benr
\textnormal{(i)}\,\,\,\cM(q,s_o(d),K)&=&\Big\{\boldsymbol\Si\,:\, \si_{ii}\le K\,\,\max_{1\le i\le d}\sum_{j=1}^{d}|\si_{ij}|^{q}\le s_0(d)\Big\}\,\,\,\textnormal{and}\nn\\
\textnormal{(ii)}\,\,\,\cU(q,s_o(d),K)&=&\Big\{\boldsymbol\Om\,:\,\boldsymbol\Om\succ 0,\,\,\|\boldsymbol\Om\|_1\le K,\,\,\max_{1\le i\le d}\sum_{j=1}^{d}|\om_{ij}|^{q}\le s_0(d)\Big\}.\nn
\eenr
Here $0\le q<1.$
\end{description}

The quantity $s_0(d)$ is allowed to depend on $d$ and thus is not and explicit restriction on sparsity. Two examples of matrices that satisfy the above restrictions are, a p-diagonal matrix that satisfies this condition with any $0\le q<1$ and $s_0(d)=K^qp.$ Second, an $AR(1)$ covariance matrix where $\si_{ij}=\rho^{|i-j|},$ which satisfies the restriction with $s_0(d)=c_0$ for some constant $c_0<\iny.$

To describe our methodology we need the following definitions. Let
\benr\lel{meanest}
\h \mu^{l}= \frac{1}{|n(l)|}\sum_{i\in n(l)}X_{ij},\qquad\,\, 1\le l\le d.
\eenr
and define a re-normalized sample covariance matrix as follows $\h{\boldsymbol\Si},$
\benr\lel{hsig}
\h\si_{lm}=\sum_{i\in n(l,m)}(X_{il}-\h\mu^{l})(X_{im}-\h\mu^{m})\Big/|n(l,m)|\quad \textnormal{and}\quad \h{\boldsymbol\Si}=\big[\h\si_{lm}\big]_{l,m=1,..,d}.
\eenr
The matrix $\h{\boldsymbol\Si}$ is an initial estimator for obtaining consistent estimators $\boldsymbol\Si$  and $\boldsymbol\Om $ of the covariance matrix and the precision matrix, respectively. Following is a key result needed for deriving the convergence rates of the estimators of $\boldsymbol\Si$ and $\boldsymbol\Om $.
\begin{lem}\lel{ecrossl}
Let $\h{\boldsymbol\Si}$ be as defined in (\r{hsig}) and assume that $\si_{ii}\le K,$ $1\le i\le d$ for some constant $K<\iny$ along with condition {\bf (A1)}. Then with probability at least $1-c_1\exp(-c_2\log d),$
\benr\lel{maxb}
\big\|\h{\boldsymbol\Si}-\boldsymbol\Si\big\|_{\iny}\le c_0\sqrt{\frac{\log d}{n}},
\eenr
for some constant $c_0<\iny.$
\end{lem}
{To appreciate this fairly innocuous result note that $\h\si_{lm},$ $1\le l,m\le d$ are defined through ${\bf X}_i,$ $1\le i\le n,$ whose distribution is in turn defined conditionally of the missing structure ${\bf M}_i.$   However, Lemma \r{ecrossl} provides an unconditional probability bound on the desired random quantity with little only a mild assumption {\bf (A1)} on the missing structure. The key to the proof of this result is the observation that $|n(l,m)|,$ $1\le l,m\le d$ is a sum of independent random variables, which allows the applicability of the Hoeffding's inequality in combination with conditional expectation arguments. The details of the proof are provided in the appendix. We now proceed with the estimation of $\boldsymbol\Si $ and $\boldsymbol\Om$.}

\subsection{{\bf Covariance Matrix}}\lel{cov}
Let $s_{\la}(x)$ be a generalized thresholding operator as defined by Rothman, Levina and Zhu (2009). We restate this definition for the convenience or the reader. A function $s_{\la}\,:\,\R\to \R$ satisfying
\benr\lel{sla}
\textnormal(i)\,\,|s_{\la}(x)|\le |x|,\quad \textnormal(ii)\,\,s_{\la}(x)=0\,\, \textnormal{for}\,\,|x|\le \la\,\,\textnormal{and}\,\, \textnormal(iii)\,\,|s_{\la}(x)-x|\le \la
\eenr
is said to be a generalised thresholding operator. In view of this definition, the covariance matrix $\boldsymbol\Si$ can be estimated by,
\benr
s_{\la}(\h{\boldsymbol\Si})=\big[s_{\la}(\h\si_{ij})\big]_{i,j=1,...,d}\nn
\eenr
The two most common examples of the thresholding operators are the hard and soft thresholding operators defined as,
\benr
s_{\la}^{H}(x)=z{\bf 1}(|x|>\la),\qquad s_{\la}^{s}(x)=sign(x)(|x|-\la)_{+},
\eenr
respectively. The soft thresholding operator can alternatively be defined as,
\benr
s_{\la}^{s}(x)=\textnormal{arg min}_{\theta} \Big\{(\theta-x)^2+\la|\theta|\Big\},\nn
\eenr
and has been studied by various authors the first of which are Donoho et. al. (1995) and Tibshirani (1996). The hard thresholding operator was first investigated by Bickel and Levina (2008) and several authors since then.  Other examples of thresholding operators include SCAD of Fan and Li (2001), the adaptive Lasso of Zuo (2008).

The following result provides the consistency of the proposed estimator.
\begin{thm}\lel{covr}
{Suppose conditions (\r{yid}),  {\bf(A1)} and {\bf (A2(i))}. Also, assume that $s_{\la}$ satisfies condition (\r{sla}).} Then, uniformly on $\cM(q, s_0(d),K)$ if $\la=K'\sqrt{\log d}/\sqrt{n}=o(1)$ for sufficiently large $K'$, then
\benr
\big\|s_{\la}(\h{\boldsymbol\Si})-\boldsymbol\Si\big\|_2=O\bigg(s_0(d)\Big(\sqrt{\frac{\log d}{n}}\Big)^{1-q}\bigg),
\eenr
with probability at least $1-c_1\exp(-c_2\log d).$
\end{thm}
{In the standard i.i.d Gaussian setting, Rothman, Levina and Zhu (2009) introduced this generalized thresholding methodology by thresholding the usual sample covariance matrix.}

\subsection{{\bf Precision Matrix}}\lel{preci}
In some problems it is of interest to estimate a precision matrix directly, for example to explore the underlying conditional independence structure via graphical models. In addition, the precision matrix under a Gaussian setup is naturally sparser in comparison to the corresponding sparse covariance matrix. Here we describe a methodology to estimate the precision matric under our structural zeros setup.

Let $\h{\boldsymbol\Om}_1$ be the solution of the following convex program,
\benr\lel{clime}
\min{\|\boldsymbol\Om\|_{1}}\quad \textnormal{subject to}\quad \big|\hat{\boldsymbol\Si}_n\boldsymbol\Om-{\bf I}\big|_{\iny}\le \la_{\Om},\quad \boldsymbol\Om\in\R^{p\times p},
\eenr
with a suitable choice of $\la_{\boldsymbol\Om}>0.$ Here ${\bf I}$ represents the identity matrix and $\h{\boldsymbol\Si}$ as defined in (\r{hsig}). Since the solution $\h{\boldsymbol\Om_1}$ may not be symmetric in general, the final estimate $\h{\boldsymbol\Om}$ is obtained by symmetrizing $\h{\boldsymbol\Om_1}=[\omega_{ij}^1]_{d\times d}$ as follows,
\benr
\h{\boldsymbol\Om}&=&(\h\om_{ij}),\quad {\textnormal{with}},\nn\\
\h\om_{ij}&=&\h\om_{ji}=\h\om_{ij}^1{\bf 1}[|\om_{ij}^1|\le|\h\om_{ji}^1|]+\h\om_{ji}^1{\bf 1}[|\om_{ij}^1|>|\h\om_{ji}^1|],\nn
\eenr
i.e., the smaller of $|\om_{ij}^1|$ and $|\om_{ji}^1|$ is chosen in the final estimate $\h{\boldsymbol\Om}.$

The following theorem provides the consistency of this methodology.
\begin{thm}\lel{fcor}
Suppose (\r{yid}) and assume condition {\bf(A1)}. If $\boldsymbol\Om  \in {\cal U}$ and $\la_{\boldsymbol\Om}=c_0\sqrt{\log d/n},$ then the following bounds hold with probability at least $1-c_1\exp(-c_2\log d),$
\benr
&\textnormal(i)&\,\,\,\|\h{\boldsymbol\Om}-\boldsymbol\Om\|_{\iny}\le O\Big(\sqrt{\frac{\log d}{n}}\Big)\nn\\
&\textnormal(ii)&\,\,\, \|\h{\boldsymbol\Om}-\boldsymbol\Om\|_{2}\le O\Big(s_0(d)\sqrt{\frac{\log d}{n}}\Big)^{1-q}\quad\textnormal{and},\nn\\
&\textnormal(iii)&\,\,\frac{1}{d}\|\h{\boldsymbol\Om}-\boldsymbol\Om\|_F^2\le O\Big(s_0(d)\sqrt{\frac{\log d}{n}}\Big)^{2-q}.\nn
\eenr
\end{thm}

{This methodology was introduced by Cai, Liu and Luo (2011) under the standard i.i.d. Gaussian setup, which is implemented using the sample covariance matrix as the initial estimate.}  The proofs for the error bounds of Theorem \r{covr} and Theorem \r{fcor} follow by deterministic arguments on the event where the inequality (\r{maxb}) holds and is thus the same as that of Rothman, Levina and Zhu and Cai, Liu and Luo respectively and are hence omitted.

\section{Simulation Study}

In this section we numerically evaluate the performance of the methodology developed in this paper. All computations were done in R. The Lasso optimizations are done by the 'glmnet' package developed by Friedman, Hastie, Simon and Tibshirani (2015) and the estimation of the precision matrix was done by the `clime' package of Cai Liu and Luo (2011). The tuning parameters $\la$ and $\la_ {\boldsymbol\Om }$ are chosen by cross validation with the loss function chosen as $\|s_{\la}(\h{\boldsymbol\Si})-\h{\boldsymbol\Si}\|_F$ and $\textnormal{Tr}(\h{\boldsymbol\Si}\h{\boldsymbol\Om}-{\bf I})^2$ respectively.

\subsection{Simulation Setup and Results}

We examine the performance of the proposed methodologies in estimating the covariance and precision matrices under two types of Gaussian graphical models, namely band and cluster structured graphs. These precision matrices are generated by the package ``fastclime’’ developed by Pang, Liu and Vanderbei (2014). For a $d$-dimensional graph, around $d/20$ band width or clusters are assumed in the two cases, respectively. The adjacency matrices of these graphs with $d=50$ are illustrated below.

\begin{figure}[]
\caption{\small{Plots of adjacency matrices of banded and cluster precision matrices respectively at d=50.}}
\vspace{1.5mm}
\begin{minipage}{.48\textwidth}
\centering
\includegraphics[scale=0.25]{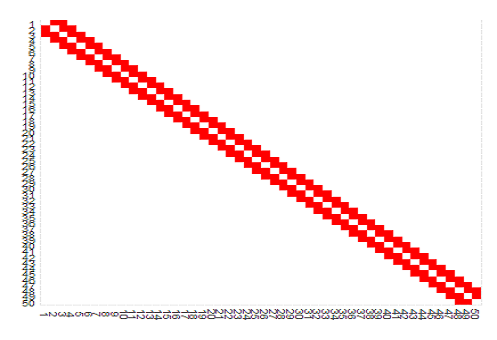}
\end{minipage}
\begin{minipage}{.48\textwidth}
\centering
\includegraphics[scale=0.25]{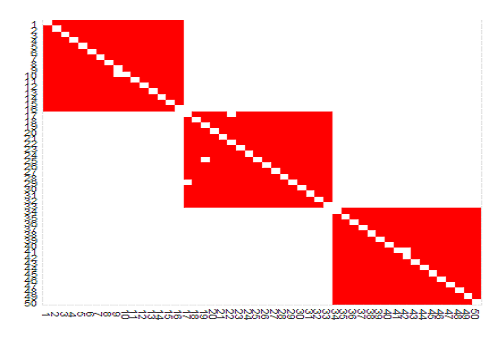}
\end{minipage}
\label{omill}
\end{figure}

The precision matrices are generated so that the corresponding covariance matrix $\boldsymbol\Si = \boldsymbol\Om^{-1}$ is a correlation matrix. For further details on the construction of these matrices see, page 5 of Pang, Liu and Vanderbei (2014).

We generate the missing structure matrix ${\bf M}_i=\big[m_{ij}\big]_{n\times d},$ as $m_{ij}\sim^{i.i.d} Bernoulli(1-\rho_j),$ $1\le i \le n,$ $1\le j\le d.$ Here $\rho_j,$ denotes the probability of $j^{th}$ component missing and they are generated by a uniform distribution between $(0, 0.75).$ For each $i$, $1\le i\le n$, the non-missing components are assumed to be normally distributed with corresponding  mean sub-vector of $\boldsymbol\mu$ and sub-block of the matrix $\boldsymbol\Si.$ Without loss of generality, the mean vector $\boldsymbol\mu $ is assumed to be a $d$-dimensional vector of zeros.

{The covariance and precision estimators derived in this paper are based on the re-normalized sample covariance matrix (\r{hsig}). In this simulation study we compare the covariance and precision estimators based on the re-normalized sample covariance matrix with those based on the usual sample covariance matrix in terms of the spectral norm loss function, i.e.  $\|\h{\boldsymbol\Si}-\boldsymbol\Si\|_2$ and $\|\h{\boldsymbol\Om}-\boldsymbol\Om\|_2$,  respectively. In the simulation experiments, the sample sizes $n$ varied from $75$ to $300$ and the dimension $d$ varied from $25$ to $175$.}

{\noi}$\bullet$ {\bf Covariance matrix}: A total of 160 independent models were generated in this study. Estimates are computed for both the hard and soft thresholding procedures described in Section \r{est}. Simulation results are illustrated in Figure \r{grafsim1} and Figure \r{grafsim2}.

{\noi}$\bullet$ {\bf Precision matrix}: A total of 112 independent models were generated in this study. Simulation results are illustrated in Figure \r{grafsim3}.

\begin{figure}[]
\caption{\small{Plots of $\|\h{\boldsymbol\Si}-\boldsymbol\Si\|_2$, against $n/\log p,$ for clustered graph model (CS) and for banded graph model (BS) for soft thresholding procedure.}}
\vspace{1.5mm}
\begin{minipage}{.48\textwidth}
\centering
{\bc \bf CS\ec}
\includegraphics[scale=0.35]{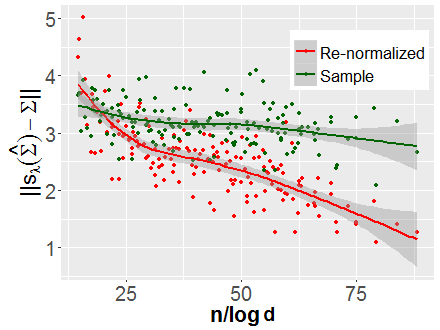}
\end{minipage}
\begin{minipage}{.48\textwidth}
\centering
{\bc \bf BS\ec }
\includegraphics[scale=0.35]{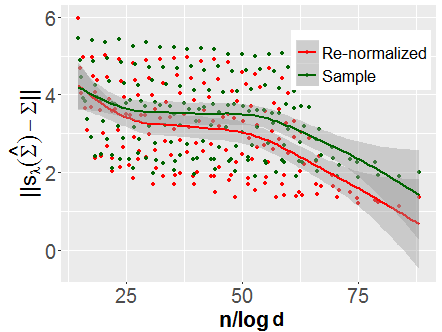}
\end{minipage}
\label{grafsim1}
\end{figure}

\begin{figure}[]
\caption{\small{Plots of $\|\h{\boldsymbol\Si}-\boldsymbol\Si\|_2$, against $n/\log p,$ for clustered graph model (CH) and for banded graph model (BH) for hard thresholding procedure.}}
\vspace{1.5mm}
\begin{minipage}{.48\textwidth}
\centering
{\bc\bf CH\ec}
\includegraphics[scale=0.35]{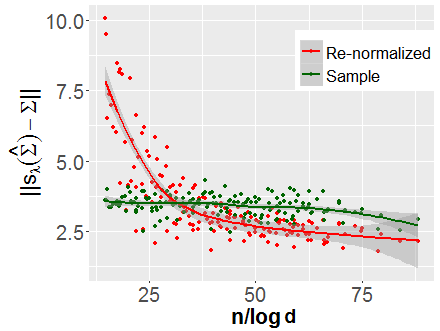}
\end{minipage}
\begin{minipage}{.48\textwidth}
\centering
{\bc\bf BH\ec}
\includegraphics[scale=0.35]{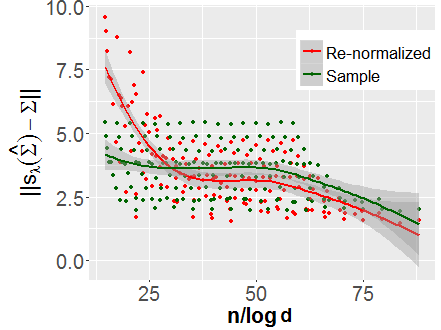}
\end{minipage}
\label{grafsim2}
\end{figure}

\begin{figure}[]
\caption{\small{Plots of $\|\h{\boldsymbol\Om}- \boldsymbol\Om \|_2$, against $n/\log p,$ for clustered graph model (CP) and for banded graph model (BP) the $\ell_1$ minimization procedure .}}
\vspace{1.5mm}
\begin{minipage}{.48\textwidth}
\centering
{\bc\bf CP\ec}
\includegraphics[scale=0.35]{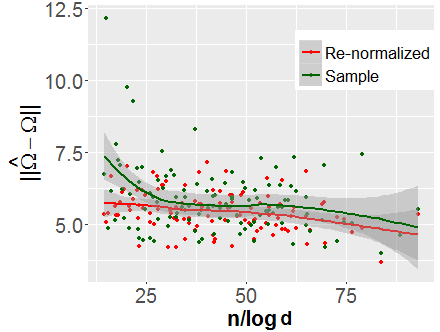}
\end{minipage}
\begin{minipage}{.48\textwidth}
\centering
{\bc\bf BP\ec}
\includegraphics[scale=0.35]{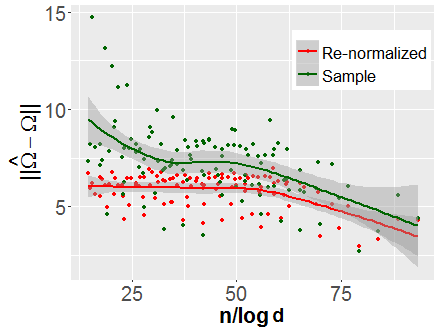}
\end{minipage}
\label{grafsim3}
\end{figure}

{Figures \r{grafsim1}, \r{grafsim2} \& \r{grafsim3} clearly illustrate consistency in estimation of both the covariance and precision matrix estimators, thus agreeing with the theoretical results. Also the proposed methodology based on the renormalized covariance almost uniformly outperforms the estimates obtained via the usual sample covariance matrix which ignores the structural zeros in data.}

\vspace{2mm}
\noi{\bf Note:} In Figures \r{grafsim1}, \r{grafsim2} \& \r{grafsim3} two colors of each dot represent the spectral norm of the estimation error in an independently generated model for two estimates being compared. To measure the average performance over the independently simulated models, non parametric regression lines and corresponding confidence bands are drawn, these are made via the Loess method with its smoothing parameter set as $0.75$.

\section{Analysis of Global Human Gut Microbiome Data}

In this section we apply the proposed methodology to analyze the global human gut microbiome data of Yatsunenko et. al. (2012). The data consists of microbial taxa counts obtained from $317$ subjects from U.S. (US), $99$ from Venezuela (VE) and $114$ from Malawi (MA). The available data can be analyzed at various levels of bacterial taxonomy. We illustrate our methodology by analyzing these data at  three levels, namely, the genus, the family and the order.  We shall generically use the term ``taxa’’ to mean either genus or family or order.

The microbiome data are measured in terms of count variables called operational taxonomic units (OTUs).  For details regarding these data one may refer to Mandal et al. (2015).  Corresponding to the $i^{th}$ sample, let ${\bf Z}_i,$ $1\le i\le n$ denote $(d+1)$ dimensional vector of counts of taxa, which are assumed to be independent over $1\le i\le n.$ Any taxon which appears in all $n$ samples is assumed to be a reference category, without loss of generality, we shall assume the $(d+1)^{th}$ taxon to be this reference taxon. We define random variables ${\bf X}_i=(X_{i1},...X_{id})^T$ where for each $1\le j\le d,$
\benr\lel{log}
 X_{ij} = \begin{cases}
            \log \Big(Z_{i,j}/Z_{i,d+1}\Big), \quad\textnormal{if}\,\, Z_{i,j}\ne 0  \\
            {\textnormal {NA}}, \hs 1in\,\,\textnormal{if}\,\, Z_{i,j}= 0
 \end{cases},
\eenr

In this definition we use `NA' to represent structural zeros since the log ratio term can also be zero valued. Also, the reference taxon is chosen as Bifidobacterium, Bifidobacteriaceae and Bifidobacteriales at the genus, family and order level respectively. As described in the Introduction, the structural zeros (represented by NA) in each observation represent taxons that are biologically absent in the specimen. Although by construction ${\bf X}_i$'s are independent over $1\le i\le n,$ however unlike Aitchison (1986), due to the structural zeros, the log ratio transformed observations cannot be assumed to be identically distributed random variables. In contrast, the distribution of ${\bf X}_i,$ $1\le i\le n$ is assumed to be as described in (\r{yid}).

Before proceeding to the analysis, we reduce the data set by retaining only those taxa that are present in at least 20\% of the samples. Although this step is not essential for our methdology, however it is done to maintain a reasonable sample size for each pair of correlations and in turn maintain reliability of estimates. In doing so, the number of taxa at the three levels reduces to 227, 99 and 52, at the genus, the family and the order levels respectively .

\vspace{2.5mm}
\noindent
{\bf Classification of subjects to geographical location}

\vspace{1.5mm}
We use the estimates of the covariance obtained by soft thresholding and precision matrices obtained in Section \r{est} to classify subjects of the above Global gut data to their respective geographical locations. For each pair of locations, a two sample t-test is performed and 10, 25 and 50 most significant components are selected. Here the t-statistic is computed only over the observed components of the log transformed observation vector. Furthermore we also perform classification among Venezuela and Malawi subjects with $d=179$ most significant components to illustrate the performance of the proposed methodology for the case $d>n$.

For each pair of locations, data is divided into a testing and training set, we randomly split $5/6^{th}$ data into training and the remaining $1/6^{th}$ in to test sets. The training set is used to estimate means of the respective populations as well as the common covariance matrix (precision matrix) using the procedures described in Section \r{est}.

Let ${\bf X}=(X_{1},..,X_{d})^{T}$ denote the $d$-dimensional observation to be classified and let $A=\{j\,\,;\,\, X_{j}\ne 0\}$ denote the collection of indices of the non-zero components of ${\bf X}$. For location $r = 1, 2$, let $\hat{\boldsymbol\mu}_{rA}$ denote the sub-vector of  $\hat{\boldsymbol\mu}_r$ and $\boldsymbol\Si_{AA}$ denote the corresponding sub-block of $\hat{\boldsymbol\Sigma}.$ Since the observation ${\bf X}$ is assumed to be conditionally Gaussian as described in \r{yid}, we can now implement the following linear discriminant function for classification.
\benr
\delta_{r}({\bf X}_A)={\bf X}_A^{T}{\h{\boldsymbol\Si}_{AA}}^{-1}\h{\boldsymbol\mu}_{rA}-\frac{1}{2}\h{\boldsymbol\mu}_{rA}^{T}{\h{\boldsymbol\Si}_{AA}}^{-1}\h{\boldsymbol\mu}_{rA}.
\eenr
We classify ${\bf X}$ into location 1 if $\delta_1 ({\bf X}_A) > \delta_2 ({\bf X}_A) $, otherwise we classify it into population 2.

Here $\h{\boldsymbol\Si}$ is the estimated covariance matrix, which can be obtained via the generalized thresholding procedure of Section 3.1 or inverting the precision matrix $\h{\boldsymbol\Om}$ obtained from Section 3.2.  Also $\h{\boldsymbol\mu}_{r}^{\star}$ is the corresponding mean sub-vector of ${\boldsymbol\mu}_r,$ $r=1,2$ which in turn is computed using the training data for each corresponding location. The observation $x$ is assigned category $1$ when $\delta_1(x^{\star})>\delta_2(x^{\star})$ otherwise assigned category $2.$

\vspace{2mm}
\noi {\it Tuning parameter:} The tuning parameters $\la$ and $\la_{\boldsymbol\Om}$ is evaluated via 5-fold cross validation within the combined training data set of the two locations being classified. Also, the loss function used to evaluate cross validation error for covariance and precision matrix estimation is chosen to be as $\|s_{\la}(\h{\boldsymbol\Si})-\h{\boldsymbol\Si}\|_F$ and $\textnormal{Tr}(\h{\boldsymbol\Si}\h{\boldsymbol\Om}-{\bf I})^2$ respectively. Also, if a pair $(l,m)$ does not occur then we set the pairwise covariance to zero.

The percentage of correctly classified observations from the test sample is computed and we repeat the above process twenty times and average the correct classification percentages over these 20 repeats as a measure of success of the procedure.

The classification results at the order, family and genus level of bacterial taxonomy are tabulated in Table \r{cl1} - Table \r{cl3}. There is a uniformly decreasing trend in the percentages of correct classification among the pairs US-MA, US-VE and VE-MA. This being possibly due to the populations of Venezuela and Malawi being microbially similar as is indicated by Figure \r{snr} of the empirical survival functions of the pairwise differences in the sample mean divided by the corresponding standard deviation, i.e. difference in the signal to noise ratio (S/N ratio).  It is clear that the difference in the S/N ratio for Malawi and Venezuela subjects is uniformly smaller than the other two pairs.

Lastly, we perform classification between Venezuela and Malawi samples at the genus level with the 179 most significant taxa using the soft thresholding method of the re-normalized sample covariance matrix. Note that the training sample size here is 178, thus allowing us to implement the procedure in the $d>n$ setup. In this case the percentage of correct classification for Venezuela, Malawi and overall are $58.5\%,$ $55.7\%$ and $57\%$ respectively.

\begin{table}[]
\centering
\caption{Classification percentages of U.S. Vs. Malawi}
\vspace{1.5mm}
\label{cl1}
\begin{tabular}{|l|c|c|c|c|c|c|c|c|c|}
\hline
\textbf{}       & \multicolumn{3}{c|}{\textbf{10 Taxa}}                                                                            & \multicolumn{3}{c|}{\textbf{25 Taxa}}                                                                            & \multicolumn{3}{c|}{\textbf{50 Taxa}}                                                                            \\ \hline
\textbf{}       & \multicolumn{1}{l|}{$\h{\boldsymbol\Om}$} & \multicolumn{1}{l|}{$s_{\la}(\h{\boldsymbol\Si})$} & \multicolumn{1}{l|}{$\h{\boldsymbol\Si}$} & \multicolumn{1}{l|}{$\h{\boldsymbol\Om}$} & \multicolumn{1}{l|}{$s_{\la}(\h{\boldsymbol\Si})$} & \multicolumn{1}{l|}{$\h{\boldsymbol\Si}$} & \multicolumn{1}{l|}{$\h{\boldsymbol\Om}$} & \multicolumn{1}{l|}{$s_{\la}(\h{\boldsymbol\Si})$} & \multicolumn{1}{l|}{$\h{\boldsymbol\Si}$} \\ \hline
\textbf{Order}  & 79.3                                & 74.5                                & 72.2                                 & 75.1                                & 71.3                                & 71.3                                 & 75.6                                & 67                                  & 65.5                                 \\ \hline
\textbf{Family} & 94.1                                & 92.2                                & 92.2                                 & 88.1                                & 92.2                                & 83.9                                 & 85.2                                & 83.1                                & 83.3                                 \\ \hline
\textbf{Genus}  & 96.6                                & 97.5                                & 97.5                                 & 93.3                                & 93.4                                & 90                                   & 92.2                                & 83.4                                & 83.8                                 \\ \hline
\end{tabular}
\end{table}

\begin{table}[]
\centering
\caption{Classification percentages of U.S. Vs. Venezuela}
\vspace{1.5mm}
\label{cl2}
\begin{tabular}{|l|c|c|c|c|c|c|c|c|c|}
\hline
\textbf{}       & \multicolumn{3}{c|}{\textbf{10 Taxa}}                                                                            & \multicolumn{3}{c|}{\textbf{25 Taxa}}                                                                            & \multicolumn{3}{c|}{\textbf{50 Taxa}}                                                                            \\ \hline
\textbf{}       & \multicolumn{1}{l|}{$\h{\boldsymbol\Om}$} & \multicolumn{1}{l|}{$s_{\la}(\h{\boldsymbol\Si})$} & \multicolumn{1}{l|}{$\h{\boldsymbol\Si}$} & \multicolumn{1}{l|}{$\h{\boldsymbol\Om}$} & \multicolumn{1}{l|}{$s_{\la}(\h{\boldsymbol\Si})$} & \multicolumn{1}{l|}{$\h{\boldsymbol\Si}$} & \multicolumn{1}{l|}{$\h{\boldsymbol\Om}$} & \multicolumn{1}{l|}{$s_{\la}(\h{\boldsymbol\Si})$} & \multicolumn{1}{l|}{$\h{\boldsymbol\Si}$} \\ \hline
\textbf{Order}  & 76.9                                & 76.1                                & 76.3                                 & 78.2                                & 74.4                                & 75.3                                 & 75.5                                & 75.2                                & 74.6                                 \\ \hline
\textbf{Family} & 76.8                                & 74.2                                & 74.9                                 & 78.1                                & 87.8                                & 87.8                                 & 75.6                                & 80.2                                & 76.6                                 \\ \hline
\textbf{Genus}  & 79.2                                & 72.7                                & 72.6                                 & 79.7                                & 90.9                                & 77.1                                 & 79.5                                & 78.5                                & 78.3                                 \\ \hline
\end{tabular}
\end{table}

\begin{table}[]
\centering
\caption{Classification percentages of Venezuela Vs. Malawi}
\vspace{1.5mm}
\label{cl3}
\begin{tabular}{|l|c|c|c|c|c|c|c|c|c|}
\hline
\textbf{}       & \multicolumn{3}{c|}{\textbf{10 Taxa}}                                                                            & \multicolumn{3}{c|}{\textbf{25 Taxa}}                                                                            & \multicolumn{3}{c|}{\textbf{50 Taxa}}                                                                            \\ \hline
\textbf{}       & \multicolumn{1}{l|}{$\h{\boldsymbol\Om}$} & \multicolumn{1}{l|}{$s_{\la}(\h{\boldsymbol\Si})$} & \multicolumn{1}{l|}{$\h{\boldsymbol\Si}$} & \multicolumn{1}{l|}{$\h{\boldsymbol\Om}$} & \multicolumn{1}{l|}{$s_{\la}(\h{\boldsymbol\Si})$} & \multicolumn{1}{l|}{$\h{\boldsymbol\Si}$} & \multicolumn{1}{l|}{$\h{\boldsymbol\Om}$} & \multicolumn{1}{l|}{$s_{\la}(\h{\boldsymbol\Si})$} & \multicolumn{1}{l|}{$\h{\boldsymbol\Si}$} \\ \hline
\textbf{Order}  & 62.2                                & 63.2                                & 63.2                                 & 60.2                                & 71.5                                & 68.4                                 & 62.0                                & 63.1                                & 58.4                                 \\ \hline
\textbf{Family} & 58.2                                & 59.5                                & 59.4                                 & 62.8                                & 62.5                                & 62.0                                 & 58.2                                & 60.7                                & 59.4                                 \\ \hline
\textbf{Genus}  & 63.1                                & 62.1                                & 64.1                                 & 61.1                                & 82.1                                & 78.5                                 & 61.1                                & 65.7                                & 59.7                                 \\ \hline
\end{tabular}
\end{table}

\begin{figure}[]
\caption{Survival  functions of SNR for different pairs}
\label{snr}
\begin{center}
\includegraphics[scale=0.35]{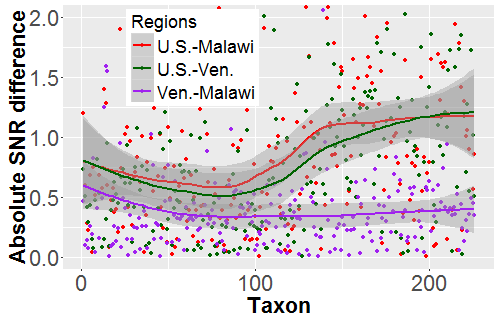}
\end{center}
\end{figure}

\section*{Acknowledgments}
Shyamal Peddada and Abhishek Kaul were supported [in part] by the Intramural Research Program of the NIH, National Institute of Environmental Health Sciences (Z01 ES101744-04). Ori Davidov was partially supported by the Israeli Science Foundation Grant No. 1256/13.

\section{Appendix}

The results to follow shall critically rely on the Hoeffding's inequality (Hoeffding (1963)).This inequality is restated below from B\:uhlmann and van de Geer (2011) for the convenience of the reader.
\begin{lem} Let $Z_1,..Z_n$ be independent r.v's with values in some space $\cL$ and let $\gamma$ be a real valued function on $\cL,$ satisfying
\benr
E\gamma(Z_i)=0,\quad |\gamma(Z_i)|\le c_i\,\,\forall\,\, i.
\eenr
Then for all $K>0,$
\benr
E\exp\big[\sum_{i=1}^n\gamma(Z_i)/K\big]\le \exp\big[\frac{\sum_{i=1}^nc_i^2}{2K^2}\big].
\eenr
\end{lem}

The \noi {\bf Proof of Lemma \r{ecrossl}} shall rely on the following two results.
\begin{lem}\lel{ecrossl1}
Let $\eta_{il}=X_{il}-\mu^{(l)},$ $1\le i\le d$ and assume conditions {\bf (A1)}, (\r{yid}) and that $\si_{ii}\le K,$ for constant $K<\iny.$ Then with probability at least $1-c_1\exp(-c_2\log d),$
\benr
\max_{1\le l,m\le d}\frac{1}{|n(l,m)|}\Big|\sum_{i\in n(l,m)}\eta_{il}\eta_{im}-\si_{lm}\Big|\le c_0\sqrt{\frac{\log d}{n}}.\nn
\eenr
\end{lem}

\vspace{3mm}
\noi {\bf Proof of Lemma \r{ecrossl1}}
Observe that
\benr\lel{ine2}
\Big|\sum_{i\in n(l,m)}\big(\eta_{il}\eta_{im}-E(\eta_{il}\eta_{im})\big)\Big|&\le& \frac{1}{4}\Big|\sum_{i \in n(l,m)}\big((\eta_{il}+\eta_{im})^2-E(\eta_{il}+\eta_{im})^2\big)\Big|\nn\\
&&+\frac{1}{4}\Big|\sum_{i \in n(l,m)}\big((\eta_{il}-\eta_{im})^2-E(\eta_{il}-\eta_{im})^2\big)\Big|\nn\\
&=& (TI)+ (TII)
\eenr
For any $1\le i\le n,$ by definition of $\eta_{il}$ and $\eta_{im},$ we have $\eta_{il}+\eta_{im},$ $1\le l,m \le d$ are conditionally Gaussian on ${\bf M}_i,$ also by elementary properties of Gaussian distributions we have
$E\big[e^{t(\eta_{iu}+\eta_{iv})^2}\Big|{\bf M}_i\big]\le c_0,$ for all $t\in \R.$ This fact can be used to show, see, for e.g. Lemma 12, Yuan (2010),
\benr\lel{expo}
E\Big[e^{t\big[(\eta_{il}+\eta_{im})^2-E(\eta_{il}+\eta_{im})^2\big]}\Big| {\bf M}_i\Big]\le e^{c_1t^2},\qquad\textnormal{for some constant}\,\,\, c_1>0
\eenr
Let ${\bf M}$ be the sigma field generated by the r.v.'s $({\bf M}_1,..,{\bf M}_n).$ Observing that $|n(l,m)|$ is entirely characterized by ${\bf M},$ we apply the exponential bound (\r{expo}) together with the Chebychev's inequality with $\la>0$ and $t=|n(l,m)|\la/2c_1,$ to obtain
\benr
P\left(\frac{1}{|n(l,m)|}\sum_{i\in n(l,m)}\big((\eta_{il}+\eta_{im})^2-E(\eta_{il}+\eta_{im})^2\big)>\la\Big| {\bf M}\right)\le \exp\big[-|n(l,m)|\la^2/4c_1\big]\nn
\eenr
Repeating this argument for the left tail and combining both we obtain,
\benr\lel{pb}
P\left(\frac{1}{|n(l,m)|}\Big|\sum_{i\in n(l,m)}\big((\eta_{il}+\eta_{im})^2-E(\eta_{il}+\eta_{im})^2\big)\Big|>\la\Big| {\bf M}\right)\le 2\exp\big[-|n(l,m)|\la^2/4c_1\big].\nn
\eenr
Now applying a trivial union bound we obtain,
\benr
P\left(\max_{1\le l,m \le d}\frac{1}{|n(l,m)|}\Big|\sum_{i\in n(l,m)}\big((\eta_{il}+\eta_{im})^2-E(\eta_{il}+\eta_{im})^2\big)\Big|>\la\Big| {\bf M}\right)\hs 1.25in\nn\\
\le \sum_{l=1}^{d}\sum_{m=1}^{d}\exp\big[-|n(l,m)|\la^2/4c_1\big]\nn
\eenr
Applying the towering and monotonic property of conditional expectation we obatin,
\benr\lel{cmain}
P\left(\max_{1\le l,m \le d}\frac{1}{|n(l,m)|}\Big|\sum_{i\in n(l,m)}\big((\eta_{il}+\eta_{im})^2-E(\eta_{il}+\eta_{im})^2\big)\Big|>\la\right)\hs 1.5in\nn\\
\le d^2 \max_{1\le l,m\le d}E\exp\big[-|n(l,m)|\la^2/4c_1\big]
\eenr
Recall the definition of $n(l,m)$ from (\r{nlm}) and observe that it can equivalently be written as,
\benr
|n(l,m)|=\sum_{i=1}^{n}I_{ilm}
\eenr
where $I_{ilm}={\bf 1}[M_{il}=1\,\,\&\,\,M_{im=1}]$  for every $1\le l,m\le d,$ where ${\bf 1}$ represents the indicator function. Note that by construction $I_{ilm}$ are independent r.v.'s over $1\le i\le n.$ Now
\benr\lel{hoeff}
\max_{1\le l\le d}E\exp\big[\frac{-|n(l,m)|\la^2}{4c_1}\big]&=&\max_{1\le l\le d}E\exp\big[-\sum_{i=1}^{n}\frac{\la^2\del(l,m)}{4c_1}\big]\exp\big[-\frac{\la^2}{4c_1}\big(|n(l,m)|-En(l,m)\big)\big]\nn\\
&\le& \exp\big[-\frac{n\la^2\del_{\min}}{4c_1}\big]\max_{1\le l\le d}E\exp\big[-\frac{\la^2}{4c_1}\big(|n(l,m)|-En(l,m)\big)\big]
\eenr
observe that $|I_i-E(I_i)|\le 2$ and apply the Hoeffdings inequality (Hoeffding (1963)) to the expected value in the r.h.s of (\r{hoeff}) to obtain,
\benr\lel{hoef}
E\exp\big[-\frac{\la^2}{4c_1}\big(|n(l,m)|-En(l,m)\big)\big]\le \exp \big[\frac{4n\la^4}{16c_1^2}\big].
\eenr
Combining (\r{hoef}) and (\r{hoeff}) with (\r{cmain}) we obtain
\benr
P\left(\max_{l,m}\frac{1}{|n(l,m)|}\sum_{i\in n(l,m)}\big((\eta_{il}+\eta_{im})^2-E(\eta_{il}+\eta_{im})^2\big)>\la\right)\hs 1.5in\nn\\
\le 2d^2\exp\big[-\frac{n\la^2\del_{\min}}{4c_1}\big]\exp\big[\frac{n\la^4}{4c_1^2}\big].\nn
\eenr
This provides a probability bound for (T1) in (\r{ine2}). Repeating the above arguments for term (TII) of (\r{ine2}) and combining it with the bound for (T1) we obtain
\benr
P\Big(\max_{l,m}\frac{1}{|n(l,m)|} \big|\sum_{i\in n(l,m)} \eta_{il}\eta_{im}-E(\eta_{il}\eta_{im})\big|\ge\la\Big)\hs 1.5in\nn\\
\le 2d^2\exp\big[-\frac{n\la^2\del_{\min}}{4c_1}\big]\exp\big[\frac{4n\la^4}{16c_1^2}\big]\nn
\eenr
Choosing $\la\ge c_0\sqrt{\frac{\log d}{n}}$ we obtain the statement of the Lemma. This completes the proof. \hfill $\Box$

\begin{rem}
In addition to the result of Lemma \r{ecrossl1}, we shall also need the following probability bound. Assuming the conditions of Lemma \r{ecrossl1} we have
\benr
\max_{1\le l\le d} \frac{1}{|n(l)|}\big|\sum_{i\in n(l)} \eta_{il}\big|\le c_0\sqrt{\frac{\log d}{n}}
\eenr
with probability at least $1-c_1\exp(-c_2\log d).$ Applying arguments similar to (\r{hoeff}) and (\r{hoef}), this result is straightforward to obtain by observing that $\frac{1}{|\sqrt{n(l)}|}\sum_{i\in n(l)}\eta_{il}$ conditioned on {\bf M} is a Gaussian r.v. with finite variance.
\end{rem}

\noi {\bf Proof of Lemma \r{ecrossl}}
Without loss of generality assume that $\mu^l=0,$ $1\le l,m\le d,$ then,
\benr\lel{iner}
|\h\sigma_{l,m}-\sigma_{l,m}|&=&\frac{1}{|n(l,m)|}\Big|\sum_{i\in n(l,m)}(X_{il}-\h\mu^l)(X_{im}-\h\mu^m)- \si_{lm}\Big|\nn\\
&\le& \frac{1}{|n(l,m)|}\Big|\sum_{i\in n(l,m)}X_{il}X_{im}-\si_{lm} \Big| + \frac{1}{|n(l,m)|}\Big|\sum_{i\in n(l,m)}\h\mu^l\h\mu^m\Big|\nn\\
&&+\frac{1}{|n(l,m)|}\Big|\sum_{i \in n(l,m)} X_{im}\h\mu^{l}\Big|+\frac{1}{|n(l,m)|}\Big|\sum_{i \in n(l,m)} X_{il}\h\mu^m\Big|\nn\\
&=& (I)+(II)+(III)+(IV),
\eenr

Term (I) of \r{iner} can be bounded by a direct application of Lemma \r{ecrossl1}. Consider Term (II),
\benr
\frac{1}{|n(l,m)|}\Big|\sum_{i\in n(l,m)}\h\mu^l\h\mu^m\Big|\le \max_{1\le l,m\le d}|\h\mu^{l}||\h\mu^{m}|\le c_0\frac{\log d}{n}
\eenr
with probability at least $1-c_1\exp(-c_2\log d).$ Lastly terms (III) and (IV) can be bounded in probability by the same arguments. Combining these bounds we obtain,
\benr
\max_{1\le l,m\le d}|\h\sigma_{l,m}-\sigma_{l,m}|\le c_0\sqrt{\frac{\log d}{n}}
\eenr
with probability at least $1-c_1\exp(-c_2\log d).$ This completes the proof of this Lemma.\hfill$\Box$

\section*{References}

\begin{enumerate}
\itemsep -0.02in

\item Aitchison, J. (1986). The Statistical Analysis of Compositional Data. London: Chapman and Hall.

\item Bickel, P. and Levina, E. (2008). Covariance Regularization by Thresholding. {\it Annals of Statistics} {\bf 36}, 2577-2604.

\item B\"{u}hlmann, P. and van de Geer, S. (2011). Statistics for High Dimensional Data, {\textit {Springer, Heidelberg}}

\item Cai, T., Liu, W. and Luo, X. (2011). A Constrained $l_1$ Minimization Approach to Sparse Precision Matrix Estimation. {\it J. of Amer. Stat. Asso.}, {\bf 106}, 594-607.

\item Clemente, J.C., Ursell, L.K., Parfrey, L.W. \& Knight, R. (2012). The Impact of the Gut Microbiota on Human Health: An Integrative View. {\it Cell} {\bf148}, 1258-1270.


\item Friedman, J., Hastie, T., Tibshirani, R. (2008). Sparse Inverse Covariance Estimation with the Graphical Lasso. {\it Biostatistics}, {\bf9}, 432--441.

\item Hoeffding, W. (1963). Probability inequalities for sums of bounded variables.   {\it J. of Amer. Stat. Asso.}, {\bf 58}, 13-30.

\item Loh, P., and Wainwright, M.J. (2012). High-dimensional regression with noisy and missing data: Provable guarantees with non-convexity. {\it Annals of Statistics} {\bf{40}}, 1637--1664.

\item Lounici, K. (2014). High Dimensional Covariance Matrix Estimation with Missing Observations. {\it Bernoulli } {\bf 20}, 1029-1058.

\item Mandal, S., Teuren, W. V., White, R. A., Eggesbo, M., Knight, R. and Peddada, S. (2015) Analysis of composition of microbiomes: a novel method for studying microbial composition. {\it Microbial Ecology in Health and Disease} {\bf 26} 1651-2235.

\item Pang, H., Liu, H., and Vanderbei, R. (2014). The fastclime Package for Linear Programming and Large-Scale Precision Matrix Estimation in R.
    {\it J. Mach. Learn. Res.}, {\bf{15}} 489-493.

\item Rothman, A., Levina, E., and Zhu, J. (2009). Generalized Thresholding of Large Covariance Matrices. {\it J. of Amer. Stat. Asso. } {\bf 104}, 177-186.

\item Yatsunenko, T., Rey, F.E., Manary, M.J., Trehan, I., Dominguez-Bello, M.G., Contreras, M., Magris, M., Hidalgo, G., Baldassano, R.N., Anokhin, A.P., Heath, A.C., Warner, B., Reeder, J., Kuczynski, J., Caporaso, J.G., Lozupone, C.A., Lauber, C., Clemente, J.C., Knights, D., Knight, R \& Gordon, J. I. (2012). Human gut microbiome viewed across age and geography. {\it Nature}, {\bf486}, 222-227.

\end{enumerate}

\end{document}